\documentclass[twocolumn,prb,showpacs]{revtex4}
\usepackage{graphicx}
\usepackage{dcolumn}
\usepackage{bm}
\usepackage{amsmath}

\begin{document}

\preprint{}
\title{A relationship between nonunitary mixed parity superconductivity and magnetism with spin-orbit coupling}
\author{Takehito Yokoyama}
\affiliation{Department of Physics, Institute of Science Tokyo, Tokyo 152-8551,
Japan 
}
\date{\today}

\begin{abstract}
We show that Hamiltonian for nonunitary mixed parity superconductivity can be recast into that for magnetism with spin-orbit coupling by the Schrieffer-Wolff transformation, indicating that nonunitary mixed parity superconductivity and magnetism with spin-orbit coupling can share the same physics.  
As demonstrations, we discuss the Dzyaloshinskii-Moriya type interactions, magnetoelectric effect, supercurrent-induced spin current, and  altermagnetism in nonunitary mixed parity superconductors. All these effects originate purely from superconductivity, without any magnetism and spin-orbit coupling.

\end{abstract}

\maketitle

The exquisite interaction between an electron's spin and its orbital motion, governed by the relativistic interaction known as spin-orbit coupling, underpins many exciting developments in contemporary condensed matter physics.
Spin-orbit coupling acts as a fundamental driving force, governing transport phenomena, dictating the electronic band structure, shaping magnetic ordering, and even giving rise to entirely new phases of matter. \cite{Manchon,Nagaosa2013,Tokura2018,Bogdanov2020,Zhai2015,Soumyanarayanan2016,Khomskii2021,Nagaosa2024}

The discovery of topological insulators has marked a paradigm shift in our understanding of phases of matter, revealing the possibility of intrinsically insulating materials that simultaneously host robust, dissipationless surface states. This remarkable behavior arises directly from the interplay between strong spin-orbit coupling and specific symmetries, highlighting the profound impact of relativistic effects on even seemingly simple materials.\cite{Hasan,QiZhang,QiHughesZhang}

Furthermore, spin-orbit coupling plays a pivotal role in the burgeoning field of spintronics, where the electric manipulation of electron spin forms the basis for information processing and storage. By influencing the spin lifetime and enabling efficient spin-current generation through effects such as the spin Hall effect, spin-orbit coupling underpins the development of spin-based transistors, magnetic random-access memories, and other novel devices with the potential to revolutionize information technology.\cite{Manchon2019,Song,Amundsen2024}

Superconductivity has been a subject of intense research for over a century. When the crystal structure possesses inversion symmetry, the parity of the gap functions is well defined: even or odd. However, recent decades have witnessed the discovery of superconductivity in noncentrosymmetric crystal structures, lacking inversion symmetry.
The absence of inversion symmetry in noncentrosymmetric superconductors profoundly impacts the behavior of superconductivity, leading to unique properties such as mixing of spin-singlet and spin-triplet (or even and odd parity) pairing states. \cite{Bauer2012,Yip2014,Samokhin2015,Smidman2017}

In addition, research on noncentrosymmetric superconductors  holds promise for advancing concepts such as topological superconductivity and Majorana fermions. Particularly, in materials with strong Rashba-type spin-orbit interaction, the realization of nontrivial topological phases is anticipated. These characteristics are expected to play a crucial role in advancing quantum computing and spintronics technologies.\cite{Alicea,Elliott2015}

In this paper, we show that Hamiltonian for nonunitary mixed parity superconductivity can be recast into that for magnetism with spin-orbit coupling by the Schrieffer-Wolff transformation, indicating that nonunitary mixed parity superconductivity and magnetism with spin-orbit coupling can manifest the same physics.  
As demonstrations, we discuss the Dzyaloshinskii-Moriya type interactions, magnetoelectric effect, supercurrent-induced spin current, and  altermagnetism in nonunitary mixed parity superconductors. All these effects are solely derived from superconductivity, without any magnetism and spin-orbit coupling.


We consider the Hamiltonian with both single and triplet pairings in a matrix form $H = {H_0} + {H_\Delta }$:
\begin{eqnarray}
{H_0} = \sum\limits_{\bf{k}} {\xi {\tau _3}{\sigma _0}} ,\; {H_\Delta } = \sum\limits_{\bf{k}} {{\tau _ + }\Delta  + {\tau _ - }{\Delta ^\dag }}
\end{eqnarray}
with
$\Delta  = {\Delta _s}({\bf{k}}){\sigma _0} + {\bf{d}}({\bf{k}}) \cdot {\bm{\sigma }},{\tau _ \pm } = \frac{1}{2}\left( {{\tau _1} \pm i{\tau _2}} \right)$ and $\xi  = \frac{{{\hbar ^2}{k^2}}}{{2m}} - {\varepsilon _F}$.
${\Delta _s}({\bf{k}})$ and ${\bf{d}}({\bf{k}})$ denote  the singlet gap function and ${\bf{d}}$-vector for triplet gap function, respectively. 
Also, $\tau $ and $\sigma$ are Pauli matrices in particle-hole and spin spaces, respectively ($\sigma _0$ is the unit matrix).
Here, we omit the tensor product. For example, ${\tau _3}{\sigma _0}$ should be interpreted as ${\tau _3} \otimes {\sigma _0}$.
Note that we use the basis such that gap functions for singlet pairing are proportional to $\sigma_0$.\cite{Ivanov2006,Yokoyama2009}
Also, notice that mixed-parity pairing can be induced by spin-orbit coupling. However, in order to clarify spin-orbit physics directly stemming from mixed parity pairing, spin-orbit coupling in the normal part of the Hamiltonian is neglected.

Let us consider $H_{\Delta}$ as a perturbation and use the Schrieffer-Wolff transformation\cite{Schrieffer1966,Bravyi2011,Landi}.
The effective Hamiltonian is then given by 
\begin{eqnarray}
H' = {H_0} + \frac{1}{2}\left[ {S,{H_\Delta }} \right] + ...
\end{eqnarray}

We take the generator of the transformation as
\begin{eqnarray}
S = \frac{1}{{2\xi }}{\tau _ + }\Delta  - \frac{1}{{2\xi }}{\tau _ - }{\Delta ^\dag }.
\end{eqnarray}
Then, we obtain the effective Hamiltonian of  the form:
\begin{eqnarray}
H'  = {H_0} + {\tau _3}\left( {{\varepsilon _{\bf{k}}}{\sigma _0} + {\bf{a}}({\bf{k}}) \cdot {\bm{\sigma }}} \right) + {\tau _0}{\bf{b}}({\bf{k}}) \cdot {\bm{\sigma }} + ...\
\end{eqnarray}
with
\begin{eqnarray}
{\varepsilon _{\bf{k}}} = \frac{{{{\left| {{\Delta _s}} \right|}^2} + {{\left| {\bf{d}} \right|}^2}}}{{2\xi }},{\bf{a}}({\bf{k}}) = \frac{1}{\xi }{\mathop{\rm Re}\nolimits} \left( {\Delta _s^*{\bf{d}}} \right),{\bf{b}}({\bf{k}}) = \frac{i}{{2\xi }}{\bf{d}} \times {{\bf{d}}^*}.
\end{eqnarray}
This perturbative treatment is justified when the gap functions are smaller than $\xi$.
A similar effective Hamiltonian has been derived in a different context.\cite{Brydon2018}

By definition, ${\bf{a}}({\bf{k}})$ and ${\bf{b}}({\bf{k}}) $ are gauge-invariant real vectors.
Since ${\Delta _s}({\bf{k}})$ and ${\bf{d}}({\bf{k}})$ are, respectively, even and odd in momentum ${\bf{k}}$,  ${\bf{a}}({\bf{k}})$ and ${\bf{b}}({\bf{k}}) $ are, respectively, even and odd in ${\bf{k}}$.
Therefore, ${\bf{a}}({\bf{k}})$ breaks inversion symmetry and plays a role of spin-orbit coupling, while ${\bf{b}}({\bf{k}}) $ breaks time-reversal symmetry and  plays a role of magnetic field.
Thus, the effective Hamiltonian in Eq.(4) can be regarded as that for magnet with spin-orbit coupling in the normal state (${\varepsilon _{\bf{k}}}$ can be considered as an effective potential). This is the central result of this paper.

This can be also expected from the general expression of the Green functions. 
The Green function is obtained as 
\begin{widetext}
\begin{eqnarray}
{G_{{\bf{k}},n}} = {\left( {i{\omega _n} - H} \right)^{ - 1}} = \frac{1}{{{A^2} - {{\left| f \right|}^2}}}\left( {\begin{array}{*{20}{c}}
{(i{\omega _n} + \xi )(A + {{\bf{f}}_ + } \cdot {\bm{\sigma }})}&{(A + {{\bf{f}}_ + } \cdot {\bm{\sigma }})\Delta }\\
{(A + {{\bf{f}}_ - } \cdot {\bm{\sigma }}){\Delta ^\dag }}&{(i{\omega _n} - \xi )(A + {{\bf{f}}_ - } \cdot {\bm{\sigma }})}
\end{array}} \right)
\end{eqnarray}
with the Matsubara frequency $\omega _n$ and 
\begin{eqnarray}
{\bf{f}}_ {\pm }({\bf{k}}) = {\bf{g}}({\bf{k}}) \pm {\bf{h}}({\bf{k}}),{\bf{g}}({\bf{k}}) = 2{\mathop{\rm Re}\nolimits} \left( {\Delta _s^*{\bf{d}}} \right),{\bf{h}}({\bf{k}}) = i{\bf{d}} \times {{\bf{d}}^*},\\
A =  - \omega_n^2 - {\xi^2} - {\left| {{\Delta _s}} \right|^2} - {\left| {\bf{d}} \right|^2}, {\left| f \right|^2} = {\left| {{{\bf{f}}_ + }} \right|^2} = {\left| {{{\bf{f}}_ - }} \right|^2} = {\left| {\bf{g}} \right|^2} + {\left| {\bf{h}} \right|^2}.
\end{eqnarray}
\end{widetext}
Note that ${{\bf{f}}_ - } \cdot {\bm{\sigma }}{\Delta ^\dag } = {\Delta ^\dag }{{\bf{f}}_ + } \cdot {\bm{\sigma }}$ holds. 
The particle component in the Green function Eq.(6) incorporates a spin-dependent term given by ${\bf{f}}_+$ which gives rise to spin-dependent physics.
 
As seen from Eqs.(5) and (7), we can expect that
 ${\bf{g}}({\bf{k}})$ and {\bf{h}}({\bf{k}}) fulfill the roles  of spin-orbit coupling and exchange field, respectively. \cite{Leggett1975}
This correspondence suggests that nonunitary mixed parity superconductivity and magnetism with spin-orbit coupling can share the same physics.  
As demonstrations, let us discuss (i)Dzyaloshinskii-Moriya type interactions, (ii)magnetoelectric  effect, (iii)supercurrent-induced spin current, and (iv)altermagnetism. All these effects can be understood from this correspondence.
Note that in the following discussions, we use the Green functions for the full Hamiltonian without the perturbative approximation. We will see that this identification is valid for all (not necessarily small)  ${\bf{g}}({\bf{k}})$ and {\bf{h}}({\bf{k}}) vectors.


(i)\textit{Dzyaloshinskii-Moriya type interactions.}  It is well known that Dzyaloshinskii-Moriya interactions stem from spin-orbit coupling.\cite{Dzyaloshinsky1958,Moriya1960} Recently, it has been predicted that Dzyaloshinskii-Moriya type interactions can arise from mixed parity superconductivity and be determined by ${\bf{g}}({\bf{k}})$.\cite{Ouassou}  
This is consistent with the above-mentioned correspondence between mixed parity superconductivity and  spin-orbit coupling.

One can distinguish between Dzyaloshinskii-Moriya type interactions arising purely from mixed parity superconductivity and those driven by spin-orbit coupling by their dependence on temperature.
Since the effects presented in this paper come from superconductivity, they vanish at the superconducting transition temperature. 
On the other hand, the Dzyaloshinskii-Moriya interactions driven by spin-orbit coupling show weak dependence on temperature around the superconducting transition temperature since they are irrelevant to superconductivity. 
Therefore, these two mechanisms are distinguishable by examining the dependence on temperature around the superconducting transition temperature.

(ii)\textit{Magnetoelectric  effect.}
It is well known that in the presence of spin-orbit coupling, supercurrent can induce spin polarization which is dubbed as the Edelstein effect.\cite{Edelstein1995}  Note that in the original paper by Edelstein, he considered a purely $s$-wave superconductor with polar point group symmetry and the Rashba type spin-orbit coupling. Subsequently, He and Law investigated the Edelstein effect in purely $s$-wave superconductors with gyrotropic point group symmetry and associated spin-orbit coupling.\cite{He2020} 
Therefore, although the systems examined in these papers lack inversion symmetry, the authors of these papers consider purely spin singlet $s$-wave superconductors with spin-orbit coupling. The Edelstein effect for mixed parity pairings has not been theoretically investigated so far. 

Now, let us show that supercurrent induced spin polarization can arise by mixed parity superconductivity even without spin-orbit coupling.
We consider a mixed parity 2D supercondutor and assume ${\bf{h}}({\bf{k}}) = 0$ since the Edelstein effect does not require magnetic field (or time reversal symmetry breaking).
We discuss the response of the spin polarization to vector potential (or supercurrent): $\left\langle {{\sigma _i}} \right\rangle  = {\chi _{ij}}{A_j}$. 
With the current operator$j_i = -\frac{e\hbar }{m}{k_i}{\tau _0}{\sigma _0}$, the coefficient ${\chi _{ij}}$ can be calculated by the Kubo formula as 
\begin{widetext}
\begin{align}
\chi_{ij} &= - \frac{e\hbar T}{mV} \sum\limits_{{\bf{k}}, n} k_j \, \text{Tr} \, \sigma_i \left( G_{{\bf{k}}, n} \right)^2 
&= - \frac{8e\hbar T}{mV} \sum\limits_{{\bf{k}}, n} k_j \frac{1}{\left( A^2 - \left| {\bf{g}} \right|^2 \right)^2} \left[ \left( \xi^2 - \omega_n^2 \right) A g_i + \mathop{\rm Re} \left( \Delta_s^* A + {\bf{g}} \cdot {\bf{d}}^* \right) \left( \Delta_s g_i + A d_i \right) \right].
\end{align}
This becomes zero for ${\bf{g}}=0$ as expected.
As an example, let us consider the $d$-vector of the form 
\begin{eqnarray}
{\bf{d}} = {\Delta _t}\left( {\begin{array}{*{20}{c}}
{\overline {{k_y}} }&{ - \overline {{k_x}} }&0
\end{array}} \right)^t
\end{eqnarray}
which may be realized in noncentrosymmetric superconductors\cite{Frigeri}. Here, $\left( {\begin{array}{*{20}{c}}
{\overline {{k_x}} }&{\overline {{k_y}} }&0
\end{array}} \right)^t$ represents the unit vector along ${\bf{k}}$.
Then, it can be calculated as 
\begin{eqnarray}
{\chi _{ij}} =  - \frac{{8e\hbar T}}{{mV}}\sum\limits_{{\bf{k}},n} {{k_j}\frac{{{\Delta _s}{\Delta _t}}}{{{{\left( {{A^2} - 4{{\left( {{\Delta _s}{\Delta _t}} \right)}^2}} \right)}^2}}}\left[ {3\omega _n^4 + 2\omega _n^2\left( {{\xi ^2} + \Delta _s^2 + \Delta _t^2} \right) - {\xi ^4} - 2{\xi ^2}\left( {\Delta _s^2 + \Delta _t^2} \right) - {{\left( {\Delta _s^2 - \Delta _t^2} \right)}^2}} \right]} {{\bf{e}}_i}
\end{eqnarray}
with ${\bf{e}} = \left( {\begin{array}{*{20}{c}}
{\overline {{k_y}} }&{ - \overline {{k_x}} }&0
\end{array}} \right)^t$.
We see that ${\chi _{xx}} =  {\chi _{yy}}=0$ holds. Now, we assume that the bands are isotropic. Then, we have ${\chi _{xy}} =  - {\chi _{yx}}$ with 

\begin{eqnarray}
{\chi _{xy}} =  - \frac{{2e\hbar T{\Delta _s}{\Delta _t}}}{{\pi m}}\sum\limits_n {\int {\frac{{{k^2}}}{{{{\left( {{A^2} - 4{{\left( {{\Delta _s}{\Delta _t}} \right)}^2}} \right)}^2}}}\left[ {3\omega _n^4 + 2\omega _n^2\left( {{\xi ^2} + \Delta _s^2 + \Delta _t^2} \right) - {\xi ^4} - 2{\xi ^2}\left( {\Delta _s^2 + \Delta _t^2} \right) - {{\left( {\Delta _s^2 - \Delta _t^2} \right)}^2}} \right]dk} } 
\end{eqnarray}
\end{widetext}
Since this is proportional to ${\Delta _s}{\Delta _t}$, we see that supercurrent induced spin polarization can stem exclusively from mixed parity superconductivity.
Near the superconducting transition temperature, we focus on the lowest orders of the gap functions and then ${\chi _{xy}}$ can be approximately calculated as 
\begin{eqnarray}
{\chi _{xy}} =  - \frac{{{\pi ^2}e\hbar {\Delta _s}{\Delta _t}}}{{4m{T^2}}}\int {\frac{{{k^2}\sinh \frac{\xi }{{2T}}}}{{\xi {{\cosh }^3}\frac{\xi }{{2T}}}}dk}\\  =  - \frac{{{\pi ^2}e{k_F}{\Delta _s}{\Delta _t}}}{{4\hbar {T^2}}}\int_{ - \infty }^\infty  {\frac{{\sinh \frac{\xi }{{2T}}}}{{\xi {{\cosh }^3}\frac{\xi }{{2T}}}}d\xi } \\
 =  - \frac{{7e{k_F}{\Delta _s}{\Delta _t}\zeta (3)}}{{2\hbar {T^2}}}.
\end{eqnarray}
Here, we set $k \approx k_F$ (the Fermi wavenumber) and ${\varepsilon _F} \to \infty$, and $\zeta (3)$ is the Riemann zeta function.
The relevant energy scale here is given by $\sqrt{|{\Delta _s}{\Delta _t}|}$ which is typically of the order of 1 meV. This is comparable to the typical energy scale of the Rashba energy splitting. Thus, we can expect a comparable size of the Edelstein like effect steming from the mixed parity superconductivity.

Let us now estimate the value of Eq.(15). For $k_F =1$ \AA$^{-1}$, $\Delta _s=\Delta _t=T=1$ meV, we have ${\chi _{xy}} \sim -6 \times 10^{15}$ (\AA V s)$^{-1}$. Then, for $A_y = 10^{-6}$ T m, we have $\left\langle {{\sigma _x}} \right\rangle  \sim -6 \times 10^{-1}$ \AA$^{-2}$. 
For comparison, the spin polarization of the electrons on the surface of topological insulators under an electric field 10$^3$ V/m is estimated as 
$\sim 5 \times 10^{-8}$ \AA$^{-2}$.\cite{Misawa2011} 

The Edelstein effect derived here can be interpreted as follows. 
As shown below (Eq.(26)), the spin distribution for a fixed energy exhibits a momentum-space texture proportional to the ${\bf{g}}$ vector which is odd in momentum, similar to that induced by Rashba-type spin-orbit coupling. Consequently, by applying an electric current, spin polarization can be achieved, analogous to the conventional Edelstein effect driven by spin-orbit coupling.

In addition to supercurrent induced spin polarization, we now investigate current induced spin polarization due to quasiparticles. 
We discuss the response of the spin polarization to an electric field: $\left\langle {{\sigma _i}} \right\rangle  = \chi {'_{ij}}{E_j}$.
The coefficient $ \chi {'_{ij}}$ can be calculated by the Kubo formula
as
\begin{eqnarray}
\chi {'_{ij}} =  - \frac{{e{\hbar ^2}}}{{4\pi mV}}\sum\limits_{\mathbf{k}} {{k_j}{\text{Tr}}{\sigma _i}G_{\mathbf{k}}^R} G_{\mathbf{k}}^A =  - \frac{{e{\hbar ^2}}}{{2\pi mV}}\sum\limits_{\mathbf{k}} {\frac{{{k_j}{{\mathbf{g}}_i}}}{{{{\left| {\mathbf{g}} \right|}^2} - {B^2}}}}
\end{eqnarray}
with $G_{\mathbf{k}}^R = {\left( {i\gamma  - H} \right)^{ - 1}},G_{\mathbf{k}}^A = {\left( { - i\gamma  - H} \right)^{ - 1}}$ and $B = \gamma _{}^2 + {\xi ^2} + {\left| {{\Delta _s}} \right|^2} + {\left| {\mathbf{d}} \right|^2}.$ Here, $\gamma$ is the damping rate  and we again assume ${\bf{h}}({\bf{k}}) = 0$.

For the $d$-vactor in Eq.(10), near the superconducting transition temperature, it can be approximately calculated as
\begin{eqnarray}
\chi {'_{xy}} = \frac{{e{\hbar ^2}{\Delta _s}{\Delta _t}}}{{2{\pi ^2}m}}\int_0^\infty  {\frac{{{k^2}}}{{{{\left| {\mathbf{g}} \right|}^2} - {B^2}}}dk}  \\ \approx \frac{{e{k_F}{\Delta _s}{\Delta _t}}}{{2{\pi ^2}}}\int_{ - {\varepsilon _F}}^\infty  {\frac{{d\xi }}{{{{(\gamma _{}^2 + {\xi ^2})}^2}}}} \\ = \frac{{e{k_F}{\Delta _s}{\Delta _t}}}{{4{\pi ^2}\gamma _{}^2}}\left( {\frac{{\gamma {\varepsilon _F}}}{{\gamma _{}^2 + \varepsilon _F^2}} + \frac{\pi }{2} + {{\tan }^{ - 1}}\frac{{{\varepsilon _F}}}{\gamma }} \right).
\end{eqnarray}
In particular, for ${\varepsilon _F} \gg \gamma $, we have 
\begin{eqnarray}
\chi {'_{xy}} \approx \frac{{e{k_F}{\Delta _s}{\Delta _t}}}{{8\pi \gamma _{}^3}}.
\end{eqnarray}
These results are analogous to the Edelstein effect in normal metals\cite{Edelstein1990,Ganichev2016} and indicate that current induced spin polarization can stem from mixed parity superconductivity, instead of spin-orbit coupling.


(iii)\textit{Supercurrent-induced spin current.} Since ${\bf{h}}$ takes on the role of magnetic field, we can expect supercurrent-induced spin current in the presence of ${\bf{h}}$. Note that spin current is time reversal even and inversion odd while charge supercurrent (or vector potential) is time reversal and inversion odd. Thus time reversal breaking field is necessary to have this effect. Supercurrent induced spin current  due to the texture of $d$-vectors in real space has been investigated in the context of superfulid $^3$He \cite{Fomin1991,Bunkov} and triplet superconductors\cite{Asano}. Here, we have reframed this effect through an alternative lens.

We assume ${\Delta _s} = 0$ since ${\bf{h}}$ does not contain ${\Delta _s}$. 
The spin current operator is given by $j_s^{ij} = \frac{\hbar }{m}{k_i}{\tau _3}{\sigma _j}$.
We consider the response of the spin supercurrent by vector potential: $\left\langle {j_s^{ij}} \right\rangle  = {\chi _{ijl}}{A_l}.$
Using the Kubo formula, $ {\chi _{ijl}}$ can be calculated as 
\begin{widetext}
\begin{eqnarray}
{\chi _{ijl}} = \frac{T}{V}\sum\limits_{{\bf{k}},n} {{\rm{Tr}}j_s^{ij}{G_{{\bf{k}},n}}{j_l}{G_{{\bf{k}},n}}}  =  - \frac{e}{V}{\left( {\frac{\hbar }{m}} \right)^2}{\sum\limits_{{\bf{k}},n} {{k_i}{k_l}{\rm{Tr}}{\tau _3}{\sigma _j}\left( {{G_{{\bf{k}},n}}} \right)} ^2}\\
 =  - \frac{eT}{V}{\left( {\frac{\hbar }{m}} \right)^2}\sum\limits_{{\bf{k}},n} {\frac{4}{{{{\left( {{A^2} - {{\left| {\bf{h}} \right|}^2}} \right)}^2}}}{k_i}{k_l}\left[ {3\omega _n^4 + 2\omega _n^2{\xi ^2} - {\xi ^4} + 2\left( {{\xi ^2} - \omega _n^2} \right){{\left| {\bf{d}} \right|}^2} - {{\left( {{\bf{d}} \cdot {\bf{d}}} \right)}^2}{{\left( {{{\bf{d}}^*} \cdot {{\bf{d}}^*}} \right)}^2}} \right]} {{\bf{h}}_j}
\end{eqnarray}
We find that when ${\bf{h}}=0$, the spin supercurrent is zero.
As an example, we consider a 2D superconductor with ${\bf{d}} = \frac{{{\Delta _t}}}{{\sqrt 2 }}\left( {\begin{array}{*{20}{c}}
1&i&0 \end{array}} \right)^t$
 and hence ${\bf{h}} = {\left| {{\Delta _t}} \right|^2}\left( {\begin{array}{*{20}{c}}
0&0&1 \end{array}} \right)^t$.
Then, we find  that ${\chi _{ijl}}$ is finite only for $i=l$ and $j=z$ with
\begin{eqnarray}
{\chi _{izi}} =  - \frac{{eT}}{\pi}{\left( {\frac{\hbar }{m}} \right)^2}{\left| {{\Delta _t}} \right|^2}\sum\limits_n {\int {\frac{{{k^3}}}{{{{\left( {{A^2} - {{\left| {{\Delta _t}} \right|}^4}} \right)}^2}}}\left[ {3\omega _n^4 + 2\omega _n^2{\xi ^2} - {\xi ^4} + 2\left( {{\xi ^2} - \omega _n^2} \right){{\left| {{\Delta _t}} \right|}^2}} \right]dk} }.
\end{eqnarray}
\end{widetext}
From this expression, we see that supercurrent induced spin current can arise from nonunitary triplet superconductivity (${\bf{h}} \ne {\mathbf{0}}$), consistent with the aforementioned correspondence. Near the superconducting transition temperature, we keep the lowest orders of the gap functions and then ${\chi _{izi}}$ becomes
\begin{eqnarray}
{\chi _{izi}} =  - \frac{{{\pi ^3}e}}{{8{T^2}}}{\left( {\frac{\hbar }{m}} \right)^2}{\left| {{\Delta _t}} \right|^2}\int {\frac{{{k^3}\sinh \frac{{\pi \xi }}{{2T}}}}{{\xi {{\cosh }^3}\frac{{\pi \xi }}{{2T}}}}dk}\\
 =  - \frac{{7{\pi }e{k^2_F}\zeta (3)}}{{4m{T^2}}}{\left| {{\Delta _t}} \right|^2}.
\end{eqnarray}
Here, we have used the same approximation to obtain Eq.(15).
Let us estimate the value of Eq.(25). 
In order to convert the  dimension of spin current into that of charge current,  we multiply Eq.(25) by $-e$.
For $k_F =1$ \AA$^{-1}$, $\hbar k_F/m =10^6$m/s, $\Delta _t=T=1$ meV and $A_i = 10^{-6}$ T m, we have $\left\langle {j_s^{iz}} \right\rangle \sim 2 \times 10^{8}$ A/m. 
For comparison, spin currents in ferromagnets under an electric field 10$^3$ V/m is estimated as $\sim 1 \times 10^{9}$ A/m$^{2}$. \cite{Amin2019}

This result can be construed in the following manner.
As shown below, the spin distribution for a fixed energy shows a momentum-space texture proportional to the ${\bf{h}}$ vector which is even in momentum. As a result, by applying an electric current, spin supercurrent can be driven, analogous to spin current generated by charge current in magnets.

(iv)\textit{Altermagnetism.} 
Since we have momentum dependent effective magnetic field, momentum-resolved spin polarization should depend on momentum as altermagnetism\cite{Ahn,Hayami,Yuan,Smejkal1,Smejkal2,Smejkal3}.
Spin polarization for fixed ${\bf{k}}$ and $\omega _n$ can be calculated as 
\begin{eqnarray}
\left\langle {{\sigma _i}({\bf{k}},\omega _n^{})} \right\rangle  = {\rm{Tr}}{\tau _0}{\sigma _i}{G_{{\bf{k}},n}} = \frac{4}{{{A^2} - {{\left| f \right|}^2}}}\left( {\xi {{\bf{h}}_i} + i\omega _n^{}{{\bf{g}}_i}} \right)
\end{eqnarray}
We see a spin texture in momentum space due to ${\bf{g}}$ and ${\bf{h}}$.
If the $d$-vector is a Rashba like one ${\bf{d}} = {\Delta _t}\left( {\begin{array}{*{20}{c}} {\overline {{k_y}} }&{ - \overline {{k_x}} }&0 \end{array}} \right)^t$ as realized in noncentrosymmetric superconductors\cite{Frigeri}, we have a Rashba like spin texture in momentum space since ${\bf{d}} || {\bf{g}}$. 

Summing this over Matsubara frequencies as
\begin{eqnarray}
\left\langle {{\sigma _i}({\bf{k}})} \right\rangle  = \frac{T}{V}\sum\limits_n {\left\langle {{\sigma _i}({\bf{k}},\omega _n^{})} \right\rangle }
\end{eqnarray}
we see that the term proportional to ${\bf{g}}$  vanishes.

As an example, let us consider  a 2D superconductor with ${\bf{d}} = {\Delta _t}\left( {\begin{array}{*{20}{c}}
{\overline {{k_x}} }&{i\overline {{k_y}} }&0
\end{array}} \right)^t$
and hence ${\bf{h}} = 2\overline {{k_x}} \overline {{k_y}} {\left| {{\Delta _t}} \right|^2}\left( {\begin{array}{*{20}{c}}
0&0&1
\end{array}} \right)^t$. 
Plugging this into the above equation, we have the spin polarization along $z$-axis:
\begin{widetext}
\begin{eqnarray}
\left\langle {{\sigma _z}({\bf{k}})} \right\rangle  = \frac{{8{{\left| {{\Delta _t}} \right|}^2}}}{V}\xi \overline {{k_x}} \overline {{k_y}} T \sum\limits_n {\frac{1}{{\left( {\omega _n^2 + {\xi ^2}} \right)\left( {\omega _n^2 + {\xi ^2} - 2{{\left| {{\Delta _t}} \right|}^2}} \right)}}} \\
 = \frac{{2{{\left| {{\Delta _t}} \right|}^2}}}{{V}}\frac{{\xi \tanh \frac{{\sqrt {\left| {{\xi ^2} - 2{{\left| {{\Delta _t}} \right|}^2}} \right|} }}{{2T}} - \sqrt {\left| {{\xi ^2} - 2{{\left| {{\Delta _t}} \right|}^2}} \right|} \tanh\frac{\xi }{{2T}}}}{{\sqrt {\left| {{\xi ^2} - 2{{\left| {{\Delta _t}} \right|}^2}} \right|} }}\overline {{k_x}} \overline {{k_y}} .
\end{eqnarray}
\end{widetext}
We find that the spin polarization shows a $d$-wave symmetry in momentum space, which is a typical form of altermagnetism.

If we consider odd frequency triplet even parity pairings instead of even frequency triplet odd parity pairings, which can be realized in ferromagnet/superconductor junctions\cite{Bergeret}, ${\bf{g}}$  becomes even in ${\bf{k}}$ and odd in $\omega _n$.
Therefore, in this case, the term proportional to ${\bf{g}}$  also contributes to the spin polarization in Eq.(27).

We have ignored strong correlation effects. This is justified when the electron-electron interaction Hamiltonian is much smaller than the single-particle Hamiltonian. The effects of the electron-electron interaction on the present results are left for future work.

In this paper, we have considered mixed parity superconductivity characterized by the Pauli matrices $\sigma$ in spin space.
Instead, if we consider multiband superconductivity by replacing the Pauli matrices $\sigma$ in spin space by the Pauli or gamma matrices in orbital space, in a similar way, we can expect orbital physics in the field of orbitronics\cite{Go2021,Jo2024,Chirolli,Mercaldo2023} even without orbital Rashba interaction.

In summary,
we have shown that Hamiltonian for nonunitary mixed parity superconductivity can be recast into that for magnetism with spin-orbit coupling by the Schrieffer-Wolff transformation, which indicates that nonunitary mixed parity superconductivity and magnetism with spin-orbit coupling can share the same physics.  
As demonstrations, we have discussed the Dzyaloshinskii-Moriya type interactions, magnetoelectric effect, supercurrent-induced spin current, and  altermagnetism in nonunitary mixed parity superconductors. All these effects originate purely from superconductivity. 

Thus far, the characteristics of spatial inversion symmetry breaking in noncentrosymmetric superconductors have been mostly incorporated as Rashba type spin-orbit coupling.
Our results reveal that broken inversion symmetry can manifest itself in a variety of physical phenomena even without Rashba type spin-orbit coupling and hence usher in new perspectives on research in noncentrosymmetric superconductors.

This work was supported by JSPS KAKENHI Grant No.~JP19K03712 and JP25K07221.

\end{document}